\documentclass[traditabstract,onecolumn,a4paper]{aa}

\usepackage{graphicx}
\usepackage{amssymb}
\begin{document}

\title{Star formation histories of early-type galaxies at z = 1.2 in cluster 
\footnote{Based on observations carried out using the ESO VLT under Program IDs 
166.A-0701,69.A-0683, 73.A-0832 and 76.A-0889} and field \footnote{Observations have 
been carried out using the Very Large Telescope at the ESO Paranal Observatory under 
Program IDs 170.A-0788, 074.A-0709 and 275.A-5060} environments}

\author{R. Gobat\inst{1} \and P. Rosati\inst{1} \and V. Strazzullo\inst{1,2} \and 
A. Rettura\inst{3} \and R. Demarco\inst{3} \and M. Nonino\inst{4}}

\institute{European Southern Observatory, Karl Schwarzschild
         Strasse 2, Garching bei Muenchen, D-85748, Germany \and  
	 National Radio Astronomy Observatory, P.O. Box O, Socorro, NM, 
	 87801, USA \and
         Department of Physics and Astronomy, Johns Hopkins University, 
         Baltimore, MD21218. USA \and
         INAF-Osservatorio Astronomico di Trieste, via G.B. Tiepolo 11, 
         34131 Trieste, Italy
}

\date{Received 7 February 2008; accepted 26 June 2008}

\abstract
{We derive the star formation histories of early-type galaxies at $z\simeq1.2$ in 
both low and high density environments. To this purpose, we compare the co-added 
spectroscopic and 8-9 band photometric 
data of 43 mass selected early-type galaxies in the massive cluster RDCS J1252.9-2927 
and the GOODS/CDF-S field with a large grid of composite stellar population models 
based on the Bruzual \& Charlot templates.
We find that the cluster early-type galaxies formed the bulk of their stars 
approximately 0.5 Gyr earlier than early-types in the field, whereas field early-types 
presumably finish forming their stellar content on a longer time scale. Such 
a difference is particularly evident at masses $\lesssim 10^{11} M_\odot$, whereas it 
becomes negligible for the most massive galaxies. While our differential analysis of the 
stellar population parameters of 
cluster and field galaxies in the same mass range convincingly shows distinct star 
formation histories, the absolute age difference remains model dependent. Using the 
star formation histories that best fit the SEDs of the red sequence galaxies in 
RDCS 1252.9-2927, we reconstruct the evolution of the cluster red sequence and find 
that it was established over 1 Gyr and is expected to dissolve by $z\approx 2$.}


\authorrunning{Gobat et al.}
\titlerunning{Star formation histories of early-type galaxies at z=1.2}

\maketitle
%

\section{Introduction}

Massive early-type galaxies are a good tracer of the early mass
assembly in the Universe. The study of their spectrophotometric and morphological
properties, especially at high redshift, over a range of environmental densities, 
can significantly constrain the different models of structure formation: 
the monolithic collapse (e.g. Eggen et al. \cite{Eggen62}), 
in which early-type galaxies result from a single burst of
star formation at high redshift, and the hierarchical model (e.g. 
Toomre \cite{Toomre77}), where
they form by the merger of low mass progenitors. This latter process
is naturally expected in a $\Lambda$CDM cosmology and predicts
different formation histories whether a galaxy is in a low-density
environment or the member of a cluster (e.g. De Lucia et al.
\cite{DeLucia06}). Indeed, the analysis of the fossil record via line-strength indices 
shows that massive early-type galaxies in local high-density environments
are at least 1.5 Gyr older than their counterparts in low-density
regions (Thomas et al. \cite{Thomas05}, S\'{a}nchez-Bl\'{a}zquez et
al. \cite{Sanchez-Blazquez06}, Clemens et al. \cite{Clemens06}), while from the 
mass-to-light ratio of cluster and field galaxies up to $z\sim1$, van Dokkum \& 
van der Marel (\cite{vanDokkum07}) find a lower value of $\sim$0.4 Gyr.
On the other hand, early-type galaxies appear to have formed at an early
time and been in place at $z \simeq 2$ (Bernardi et al.
\cite{Bernardi98}, van Dokkum et al. \cite{vanDokkum01}), with little
star formation happening ever since.  This suggests that studying the
star formation history of early-type galaxies at $z > 1$ allows one to
place stronger constraints on structure formation models than at low
redshift, where any difference has been smoothed out by billions of
years of mostly passive evolution and occasional mergers. In such a case, 
much sparser data are available, as few massive galaxy clusters have been
observed so far at $z>1$. One of these, RDCS J1252.9-2927 (Rosati et
al. \cite{Rosati04}) at $z=1.24$, has had an extensive multi-wavelength 
spectroscopic coverage. In this paper, we use the spectrophotometric data of
early-type galaxies in RDCS J1252.9-2927 to reconstruct their general
star formation history and compare it with early-type galaxies from
the GOODS/CDF-S at similar redshift. An independent analysis of the same 
data sets, also including morphological properties and far-UV rest-frame 
photometry is presented in Rettura et al. (\cite{Rettura08}). This paper is 
structured as follows. In Sect. 2, we describe our data and sample selection; 
in Sect. 3 we present our methodology and in Sect. 4 we describe and
discuss the results of our analysis.  We assume a $\Lambda$CDM
cosmology with $\Omega_m=0.3$, $\Omega_{\Lambda}=0.7$ and $H_0=70$ km
s$^{-1}$ Mpc$^{-1}$. All magnitudes in this paper are given in the AB
system (Oke \cite{Oke74}), unless stated otherwise.


\section{Data and sample selection}

This work is based on photometric and spectroscopic data from the GOODS/CDF-S 
(hereafter GOODS, Giavalisco et al. \cite{Giavalisco04}) field and 
the field around the cluster RDCS J1252.9-2927 (hereafter RDCS 1252; Rosati et al. 
\cite{Rosati04}) at $z = 1.237$ (Demarco et al. \cite{Demarco07}).

The data for the GOODS field comprises observations in 8 bands in the 0.4-5 
$\mu$m range, namely the F435W, F606W, F775W and F850LP (hereafter $b_{435}$, 
$v_{606}$, $i_{775}$ and $z_{850}$ respectively) passbands with \emph{HST}/ACS, 
the J and K$_{s}$ bands with \emph{VLT}/ISAAC (Retzlaff et al., in prep.) and at 
3.6 and 4.5 $\mu$m with \emph{Spitzer}/IRAC.In addition, we use spectra taken using 
\emph{VLT}/FORS2 with the 300I grism, which provides a resolution of 
about 13$\AA$ at 8600 $\AA$ (Vanzella et al. \cite{Vanzella06}, \cite{Vanzella08}).

For RDCS 1252, we used imaging in the B, V and R bands with
\emph{VLT}/FORS2, $i_{775}$ and $z_{850}$ with \emph{HST}/ACS
(Blakeslee et al. \cite{Blakeslee03}), J$_{s}$ and K$_{s}$ with
\emph{VLT}/ISAAC (Lidman et al. \cite{Lidman04}) and at 3.6 and 4.5
$\mu$m with \emph{Spitzer}/IRAC. An extensive spectroscopic campaign
of this cluster was carried out with \emph{VLT}/FORS2 using the 300I
grism (Demarco et al. \cite{Demarco07}).

As a result, the two datasets used in our analysis have similar
quality and wavelength coverage, thus yielding homogeneous
spectrophotometric properties of galaxies and allowing a single
selection criterion for both samples. Specifically, the availability
of 8 to 9 passbands allows the estimate of reliable photometric
stellar masses (Rettura et al. \cite{Rettura06}). This work is based on two 
samples of early-type galaxies with
available spectroscopy and covering the same mass range. One is
extracted from the cluster spectroscopic sample and compared with a
field sample drawn from the GOODS spectroscopic survey of galaxies in
a small redshift bin around the cluster redshift.  For all galaxies
we have photometric stellar masses derived by SED fitting with
composite stellar population models (Rettura et al.
\cite{Rettura06}). These were computed assuming solar metallicity and
a Salpeter (\cite{Salp55}) initial mass function, 
using the Bruzual \& Charlot (\cite{BC03}, hereafter BC03) code.

The near-IR depth of both samples allows us to define complete
mass-selected samples. The GOODS and RDCS 1252 $K_s$-band images are
photometrically complete down to $K_s = 24$. At $z \simeq
1.2$, the $K_s$-band photometry is a very good proxy of the stellar
mass (Gavazzi et al. \cite{Gavazzi96}), 
with $10^{10}M_{\odot}$ corresponding to $K_s \simeq 23$
(Strazzullo et al. \cite{Strazzullo06}), which we take as a reliable
mass completeness limit.  On the other hand, the spectroscopic
follow-up work is limited to $K_s \simeq 22$ for early-type galaxies
in both samples, corresponding to $R_{Vega} \simeq 25$ and $M_{\star}
\simeq 3\times10^{10}M_{\odot}$. Therefore, we limit our analysis to
photometric stellar masses greater than
$M_{lim}=5\times10^{10}M_{\odot}$.

The selection of early-type galaxies in the cluster is apparent from
the well-defined red sequence (Blakeslee et al. \cite{Blakeslee03})
which clearly lies above $i_{775}-z_{850} = 0.8$, which thus
separates the population of blue star-forming galaxies at the cluster
redshift. We have spectra for 22 of the 38 galaxies on the red sequence
with $M_{\star} \geq M_{lim}$.
For the corresponding GOODS field sample, we select galaxies with
$M_{\star} \geq M_{lim}$, in the redshift range $z = 1.237 \pm 0.15$
and adopt the same $i_{775}-z_{850} \geq 0.8$ cut used in
RDCS 1252. This criterion yields 21 early-type galaxies with FORS2
spectra.
We verify that the spectra do not show signs of ongoing star formation, i.e.
that no [OII]$\lambda$3727 emission line is detectable (we 
assume that the galaxies in our samples are dust-free; see Rettura et al. 
(\cite{Rettura06})). We note that the vast 
majority of the galaxies in both samples have E/S0 morphologies following the 
classification scheme used by Blakeslee et al. (\cite{Blakeslee03}). Of the 18
galaxies in our GOODS sample found in the GEMS catalog (H\"{a}ussler et al. 
\cite{Haeussler07}), all have $n\geq3$.

\begin{figure}
\centering
\includegraphics[scale=0.5]{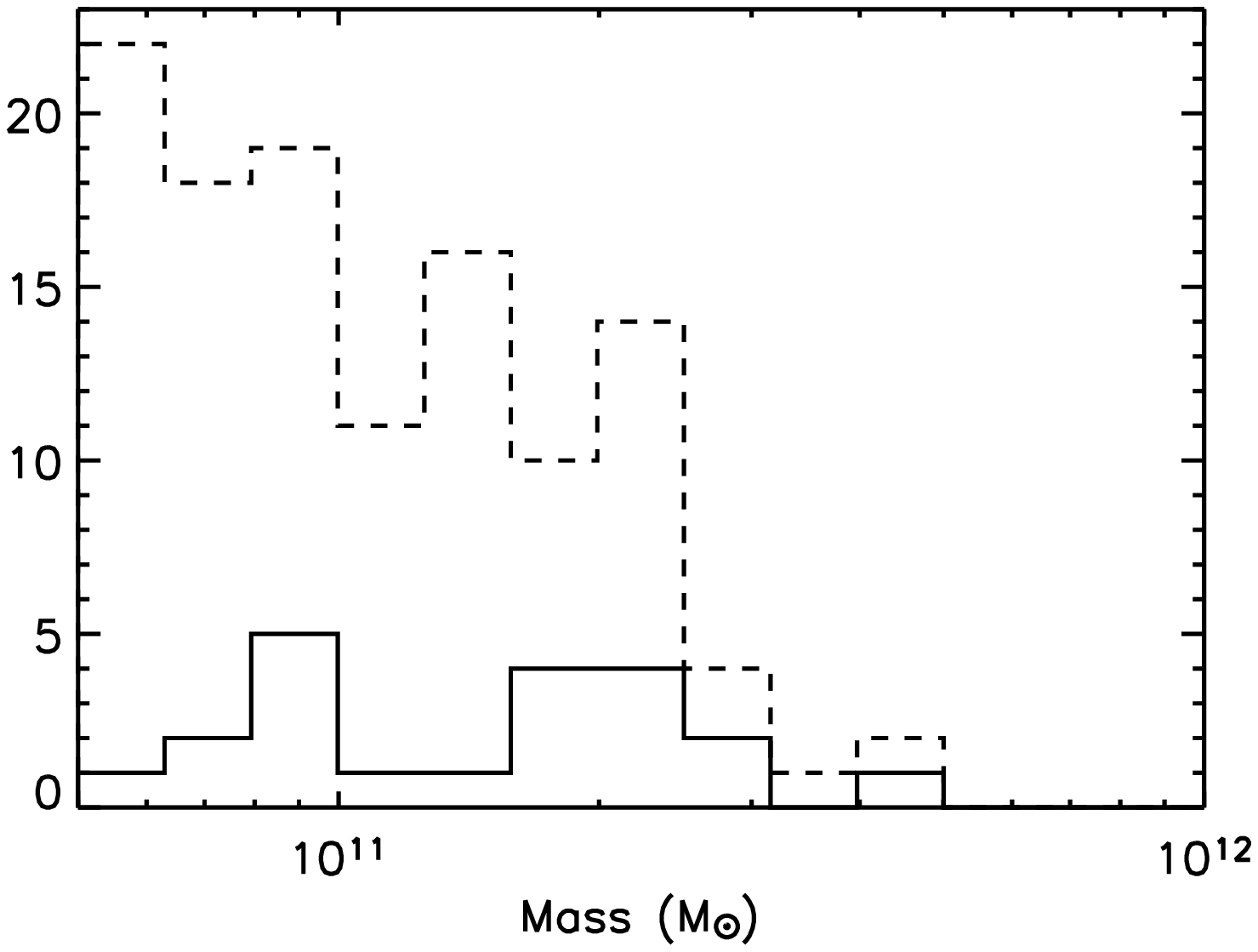}
\includegraphics[scale=0.5]{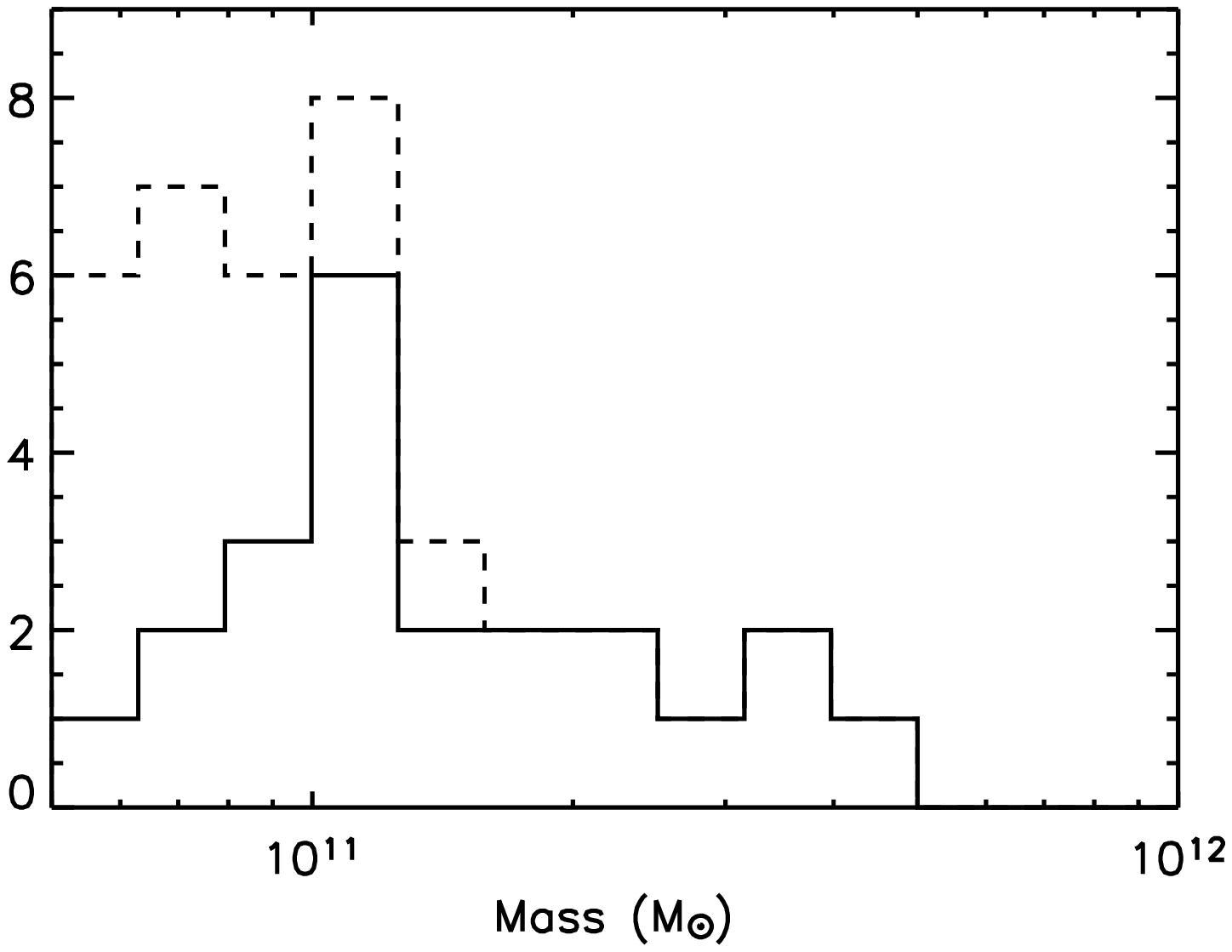}
\caption{Number of galaxies in the GOODS (\emph{left}) and RDCS 1252 (\emph{right}) 
samples as a function of the photometric stellar mass. The dashed line shows the 
photometric sample and the solid line the spectroscopic sample.}
\label{fig1}
\end{figure}

An important aspect to be addressed is the relative spectroscopic completeness as a 
function of mass of the GOODS and RDCS 1252 samples, which should be considered when 
interpreting the results from our comparative spectrophotometric analysis. This can 
be quantified with accurate photometric redshifts available for our samples, which 
were derived using the Coleman et al. (\cite{Coleman80}) templates and the BPZ code 
(Ben\'{i}tez \cite{Benitez00}), as described in Toft et al. (\cite{Toft04}). Using 
photometric redshifts, we find 117 GOODS galaxies in the given redshift bin, with 
$i_{775}-z_{850} \geq 0.8$ and $M_{\star} \geq M_{lim}$.

In Fig. \ref{fig1}, we compare the photometric mass distributions of our color- 
and mass-selected galaxies in GOODS and RDCS 1252 to the mass distribution of the 
spectroscopically observed subsamples. It is immediately apparent that the 
spectroscopic follow-up work was more extensive in RDCS 1252 than in GOODS; as a result 
the GOODS spectroscopic sample is more incomplete at the low mass end than the RDCS 1252 
sample (table \ref{tab1}). We will return to this point when discussing our results.

\begin{table}
\centering
\begin{tabular}{ccccc}
\hline
\hline
sample&\multicolumn{4}{c}{Lower mass cutoff ($\times10^{10}M_{\odot}$)}\\
&$5$&$9$&$15$&$23$\\
\hline
GOODS&18\%&25\%&35\%&67\%\\
RDCS 1252&59\%&83\%&100\%&100\%\\
\hline
\end{tabular}
\caption{Cumulative spectroscopic completeness of the GOODS and RDCS 1252 samples.}
\label{tab1}
\end{table}


\section{Spectrophotometric fitting method}
\label{sec:method}

We compare our spectrophotometric data to composite stellar population
(CSP) synthesis models generated with the GALAXEV population synthesis
code (Bruzual \& Charlot \cite{BC03}) using a $\chi^2$
goodness-of-fit test. We choose the BC03 models for their
high-resolution templates which allow us to fit both broad-band
photometry and spectra. Because of the low signal-to-noise ratio of
the available FORS2 spectra (ranging from 5 to 6 at 8000 \AA), we
resort to stacking spectra of galaxies within the aforementioned
stellar mass range separately for our cluster and field
samples. Consequently, the photometric points are also co-added in the
combined fit.  We assume a delayed exponential star formation history
(SFH), which is similar to the one proposed by Sandage
(\cite{Sandage86}) and is more realistic than a simple exponentially
declining SFH (Gavazzi et al. \cite{Gavazzi02}), parametrized by a
time-scale $\tau$. Since we use both broad-band SEDs and spectroscopic
features, we allow for more complex star formation histories. We expand the
grid of models by adding secondary episodes of star formation after
the main event parametrized by an instantaneous burst at time
$t_{burst} > \tau$ of amplitude $A$, so that the star formation rate
is expressed as :

\begin{equation}\Psi(t)=\tau^{-2}t e^{\frac{{-t}}{\tau}}+A\delta(t-t_{burst})
\,.
\label{eq1}
\end{equation}

\begin{figure}
\centering
\includegraphics[scale=0.5]{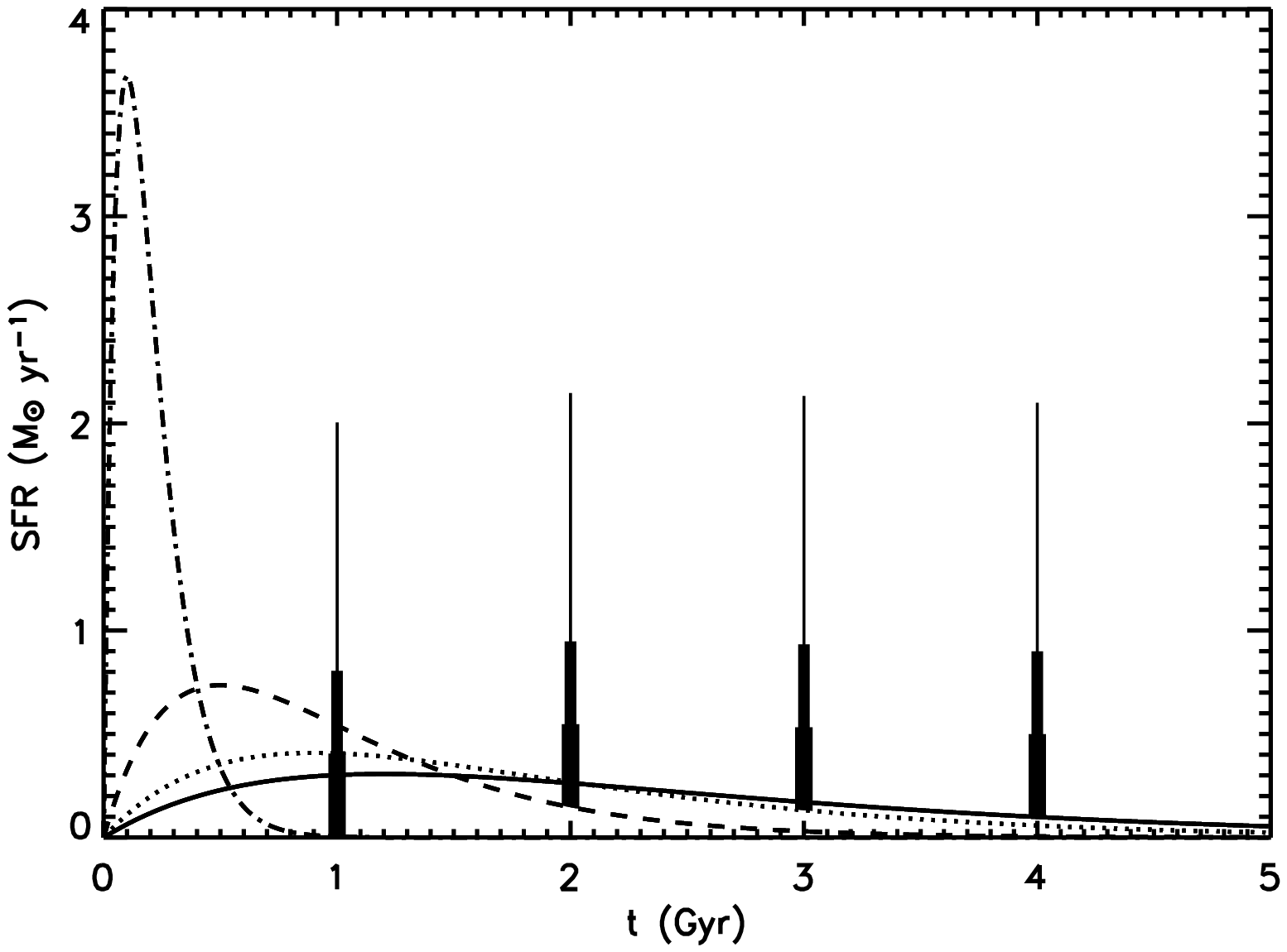}
\includegraphics[scale=0.5]{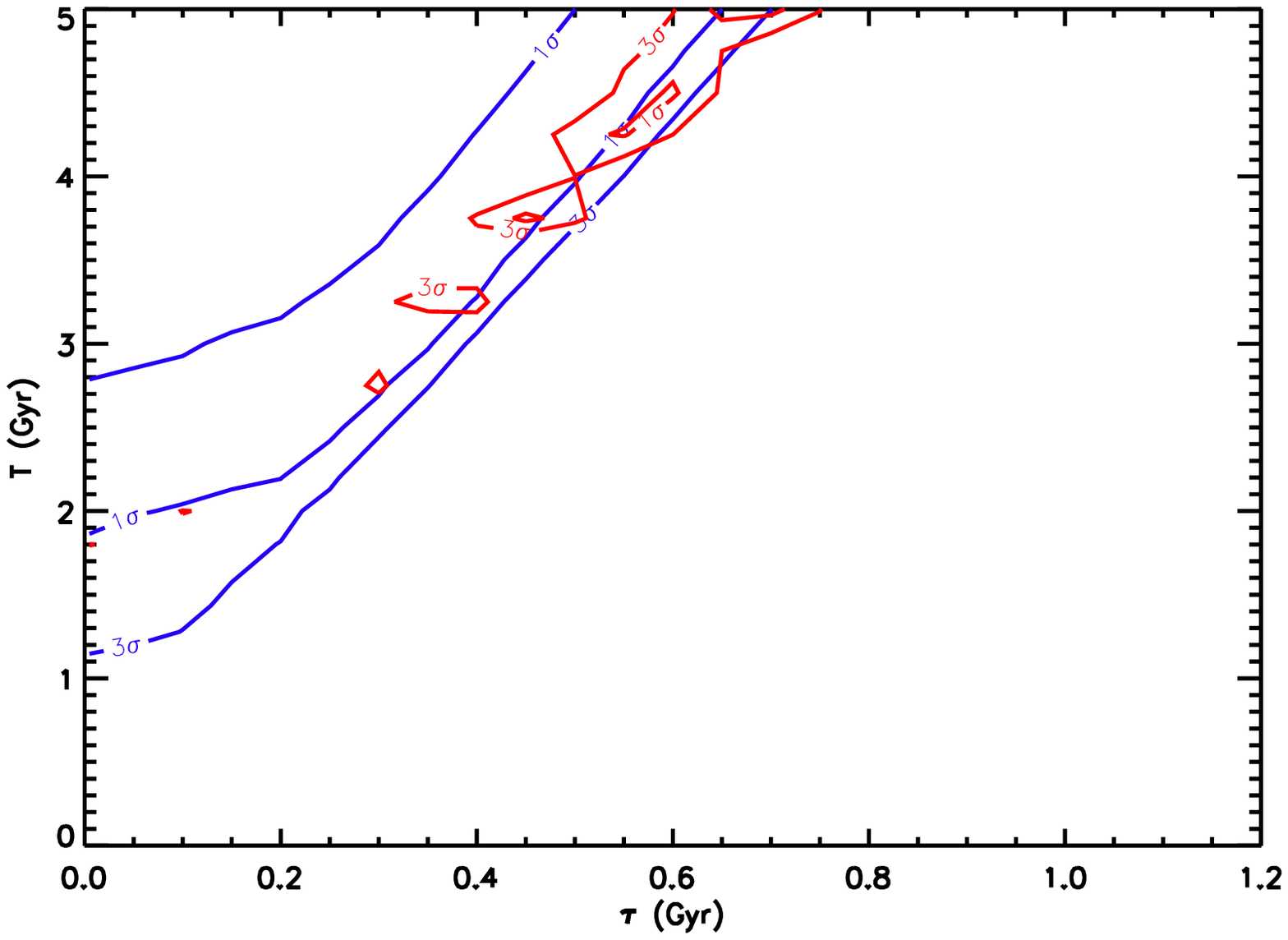}
\caption{\emph{Left panel}: range of star formation histories spanned by the grid 
of BC models (the secondary bursts are not to scale with respect to the delayed 
exponential). 
 \emph{Right panel}: confidence contours (1$\sigma$ and 3$\sigma$) obtained by
  fitting the co-added photometry (\emph{blue}) and
  spectroscopic data (\emph{red}) from the GOODS sample with a
  delayed exponential SFH without secondary burst.}
\label{fig2}
\end{figure}

We consider age values, i.e. the time $T$ after the onset of star
formation, ranging from 200 Myr to 5 Gyr in increments of $\sim$ 250
Myr. The time-scale $\tau$ ranges from 0 (corresponding to a simple
stellar population) to 1 Gyr in increments of 0.1 Gyr. In the case of a
secondary burst, $t_{burst}$ ranges from 1 to 4 Gyr in increments of
1 Gyr with amplitudes $A$ of 0.1, 0.2 and 0.5, which correspond to
1/11, 1/6 and 1/3 of the final stellar mass respectively. Thus, the
grid \{$T,\tau,t_{burst},A$\} contains
$20\times13\times4\times3+13\times20=3380$ models. All models are
computed at solar metallicity and are dust-free. We will comment on this
assumption later. In Fig. \ref{fig2}, left, we show a sample of our grid 
of star formation histories.

We first use this grid of models to fit the stacked
SEDs, covering rest-frame wavelengths from 2000 \AA~ to 2 $\mu$m,
which gives us a sub-grid of acceptable models within the confidence
interval defined by $\chi^2 \leq \chi^2_0 + 16.25$ (where $\chi^2_0$ is the 
minimum $\chi^2$ solution) which corresponds to 3$\sigma$ for the 5 fitting 
parameters: age,$\tau$, $t_{burst}$, $A$ and stellar mass.  Among this 
subsample of models, we select those which best fit the stacked spectral data,
adopting the same 3$\sigma$ confidence level. These solutions will be referred to 
as the ``best fit models'' in the following.

\begin{figure}
\centering
\includegraphics[scale=0.5]{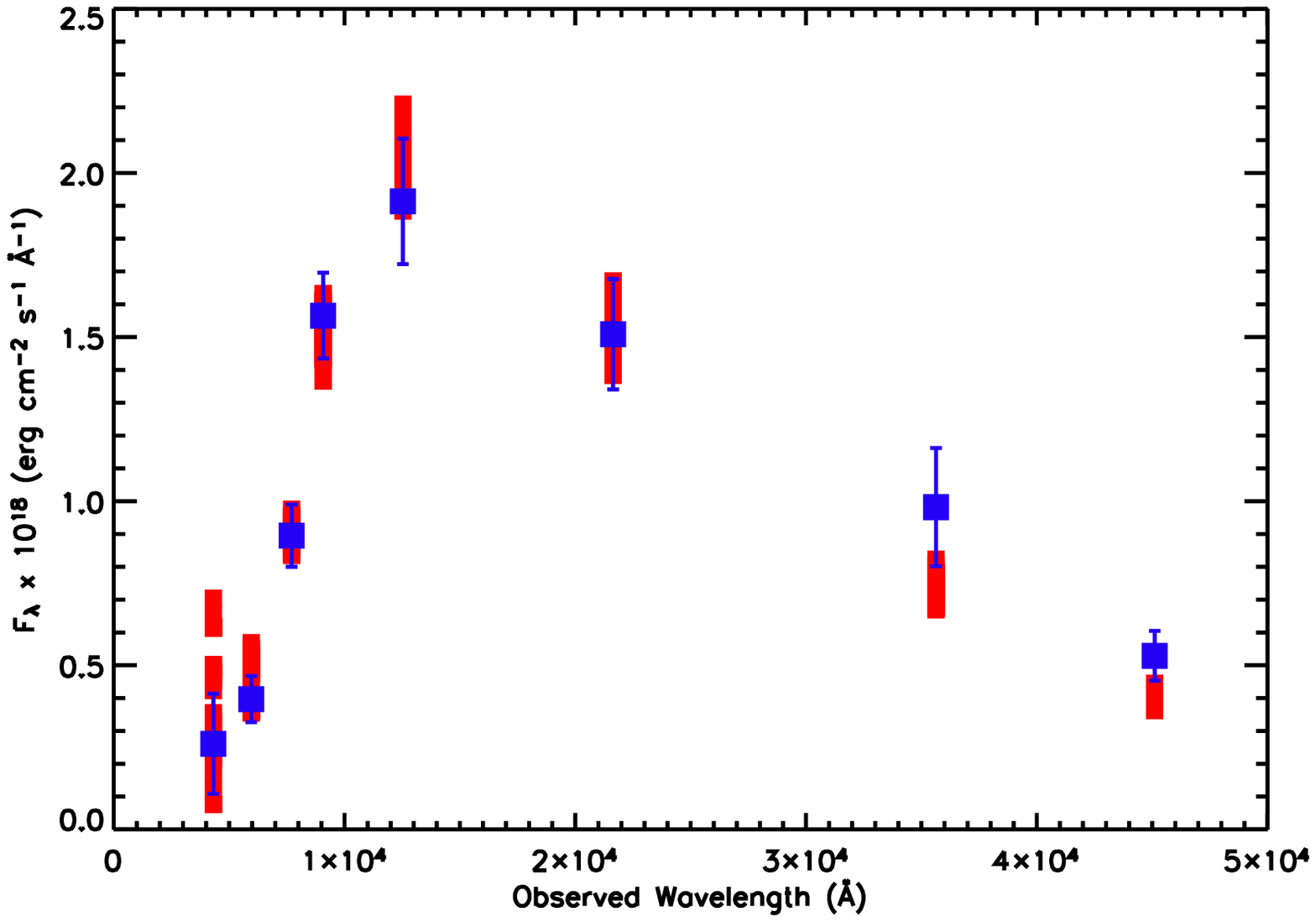}
\includegraphics[scale=0.5]{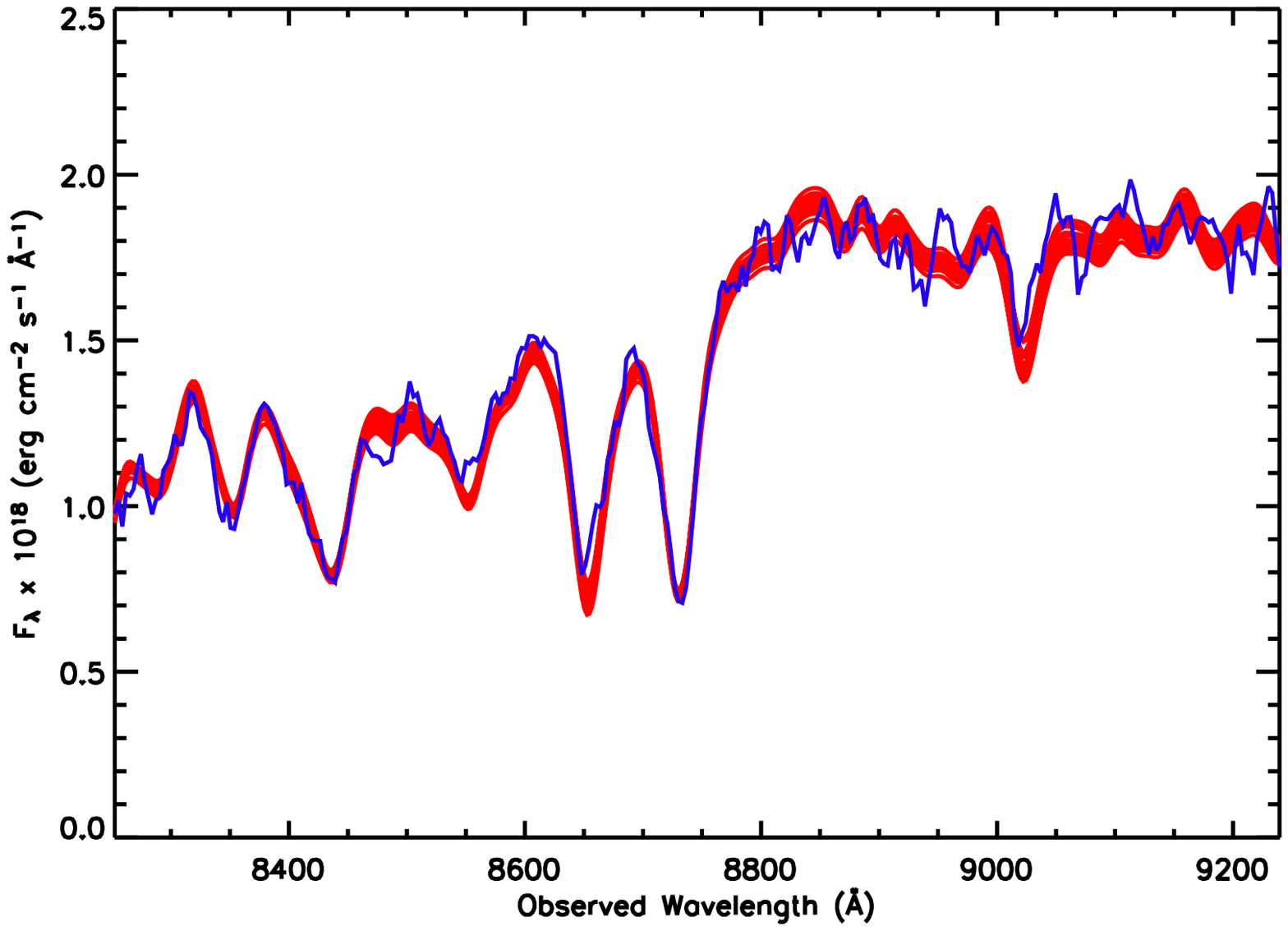}
\caption{Results from the fit of the GOODS spectrophotometric data
  with a delayed exponential SFH without secondary burst. \emph{Left
    panel}: averaged observed (\emph{blue} data points) and model
  SEDs (\emph{red}); \emph{Right panel}: stacked observed
  spectrum (\emph{blue}) and models within 3$\sigma$ confidence
  intervals.}
\label{fig3}
\end{figure}

In Fig. \ref{fig2}, right, we show the projection of the 3$\sigma$ confidence 
levels of the fit on the stacked SEDs and spectra of the GOODS sample in the 
\{$T,\tau,t_{burst}=0 {\rm (no burst)},A=0$\} plane. In Fig. \ref{fig3}, we 
show the SEDs and spectra of the best fitting models, with the stacked 
spectrophotometric data.

The model spectra are smoothed to match the resolution of the GOODS
and RDCS 1252 samples and cropped to a 500 \AA\ interval centered on
the 4000 \AA\ break, which roughly corresponds to the high S/N, well
flux-calibrated part of the spectra in our sample. 
This region also covers
absorption features characteristic of young (e.g. H$_{\delta}$) and
old (e.g. CaII H\&K) stellar populations. It includes the 
4000 \AA\ break, a good age indicator 
with mild metallicity sensitivity (e.g. Poggianti \& Barbaro 
\cite{Poggianti97}), and
therefore removes part of the degeneracy inherent to our grid of star
formation histories. By using this relatively narrow wavelength
interval, possible distortions due to uncertain flux calibration on
the red end of the spectra (Demarco et al. \cite{Demarco07}) are also
minimized.

The BC03 models offer the choice of two initial mass functions (IMFs),
Salpeter or Chabrier (2003).  The choice of the IMF has little effect
on the shape of the composite spectrum for the considered age range (t
$\leq$ 5 Gyr), as the Salpeter and Chabrier IMFs are nearly identical
above 1 $M_{\odot}$. We assume a Salpeter IMF for consistency with the
photometric masses described in the previous section. The lower and upper
mass cut offs are 0.1 and 100 $M_{\odot}$ respectively (Bruzual 
\& Charlot \cite{BC03}).

Photometric errors are described in Rettura et al. (\cite{Rettura06}). Spectroscopic 
errors are based on the evaluation of the
signal-to-noise ratio at 4100 $\AA$ rest-frame from the residuals
after fitting the $H_{\delta}$ absorption line of the co-added
spectra. The total response of the detector is then normalized to this
value to obtain a noise estimate as a function of wavelength.

The stacked spectrum of the GOODS sample has a S/N of approximately 44 and an 
equivalent exposure time of $t_{exp}=93 h$. For the RDCS 1252 sample, the corresponding 
values are S/N$\sim$20 and $t_{exp}=74 h$.


\section{Results}

To characterize the star formation history of the best fit models in
our field and cluster samples, we define the final formation time $t_{fin}$ of a
given model as the time, after the onset of star formation, at which
99\% of the final stellar mass has been formed. The corresponding
look-back time since $z=1.24$ to the final formation redshift $z_{fin}$ is
$T-t_{fin}$.

\begin{figure}
\centering
\includegraphics[scale=0.5]{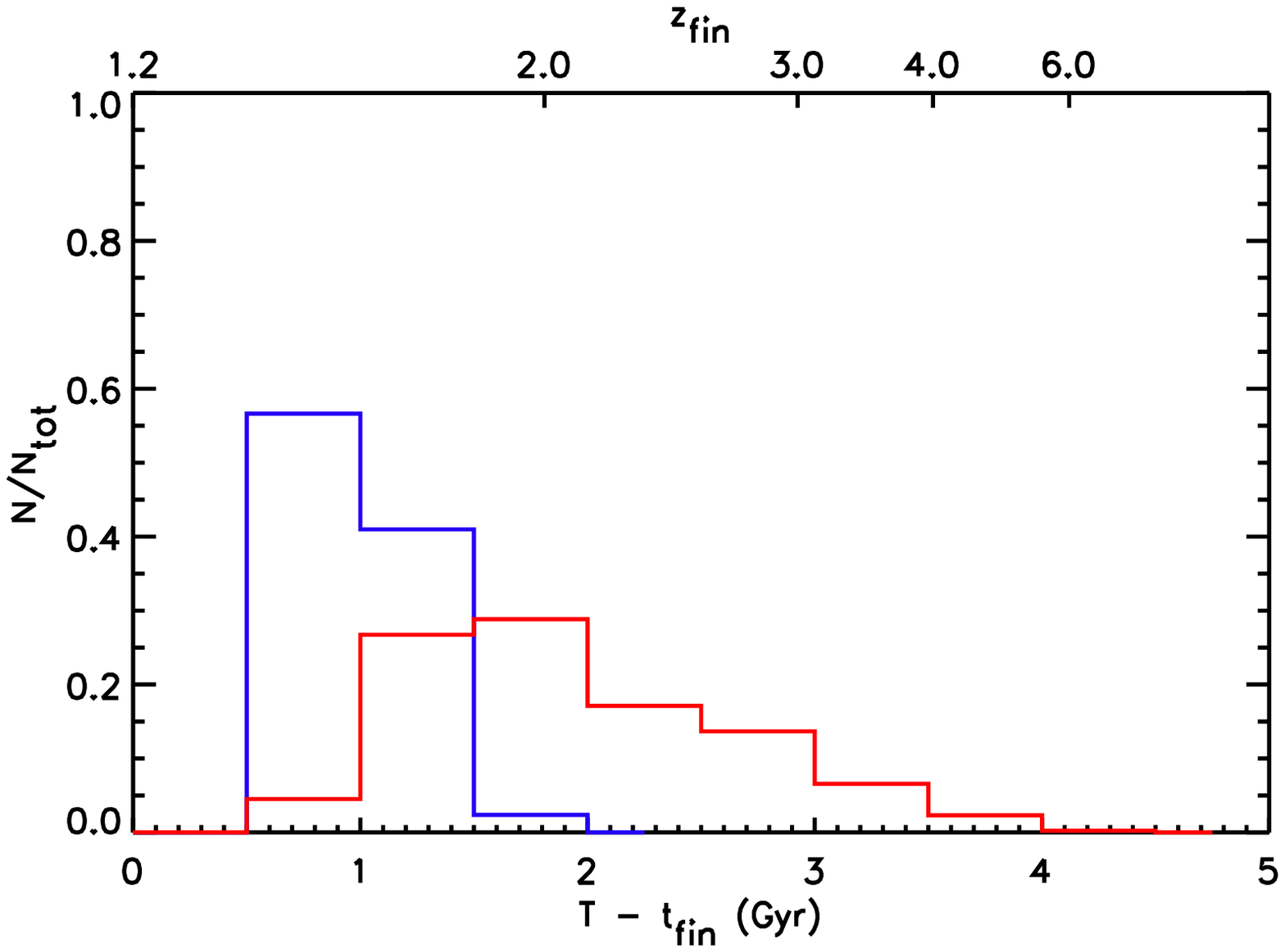}
\includegraphics[scale=0.5]{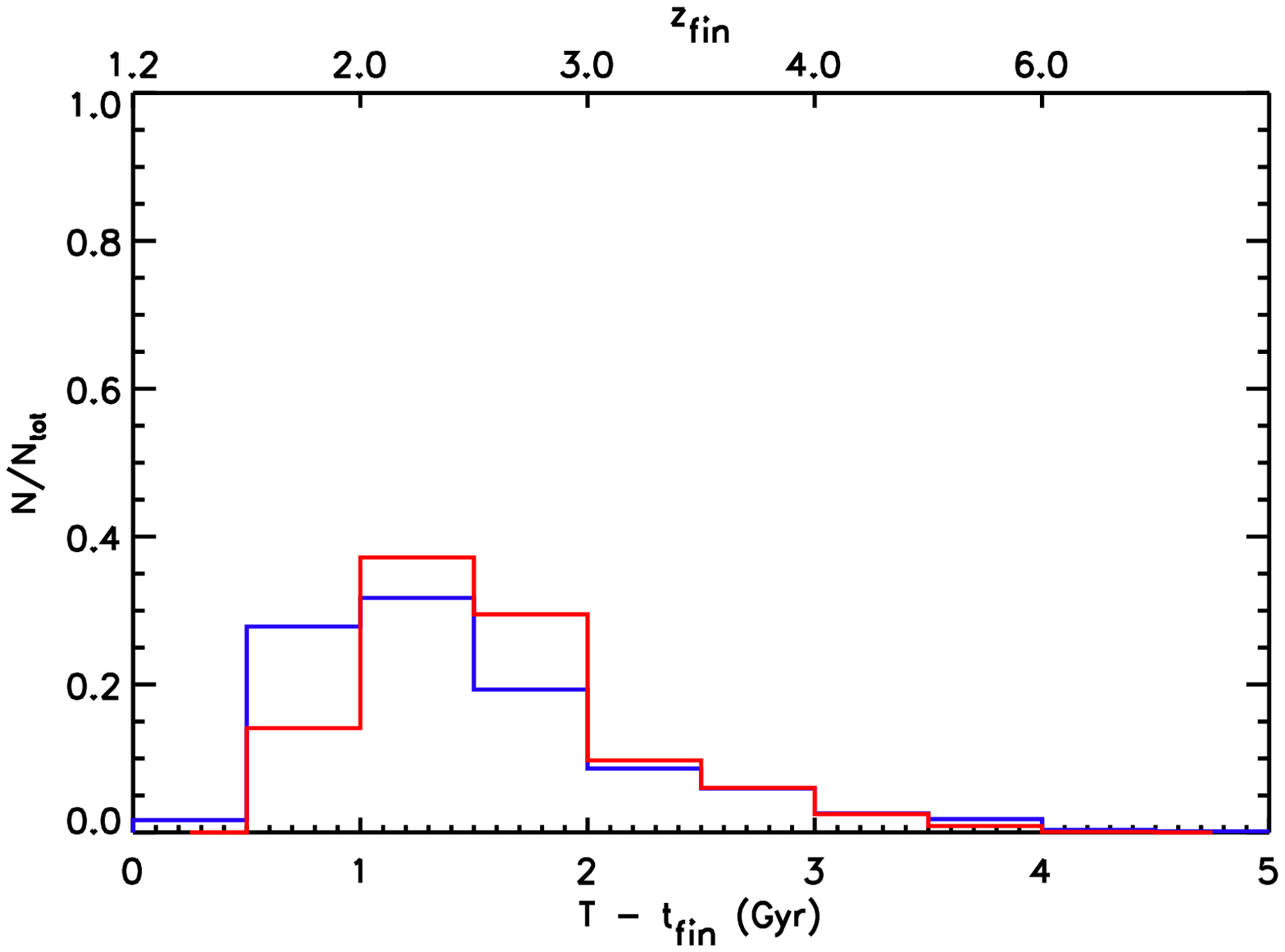}
\caption{Fraction of best fit models to the spectrophotometric (\emph{left panel}) 
and photometric only (\emph{right panel}) data, within the 3$\sigma$ contours, as a 
function of final formation redshift $z_{fin}$ and look-back time since $z=1.23$, 
$T-t_{fin}$. The blue histogram corresponds to the GOODS field sample and the 
red one to the RDCS 1252 cluster sample.}
\label{fig4}
\end{figure}

\begin{figure}
\centering
\includegraphics[scale=0.5]{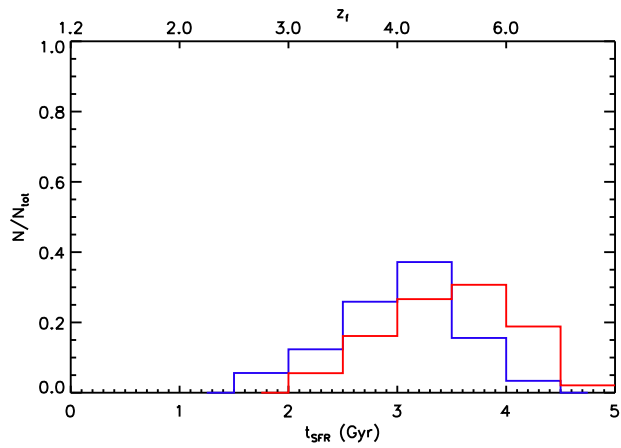}
\caption{Fraction of best fit models for the field (\emph{blue}) and cluster 
(\emph{red}) samples, as in Fig. \ref{fig4}, as a function of the star-formation 
weighted age $t_{SFR}=\int_0^T dt'(T-t')\Psi(t')/\int_0^T dt'\Psi(t')$.}
\label{fig5}
\end{figure}

\begin{figure}
\centering
\includegraphics[scale=0.5]{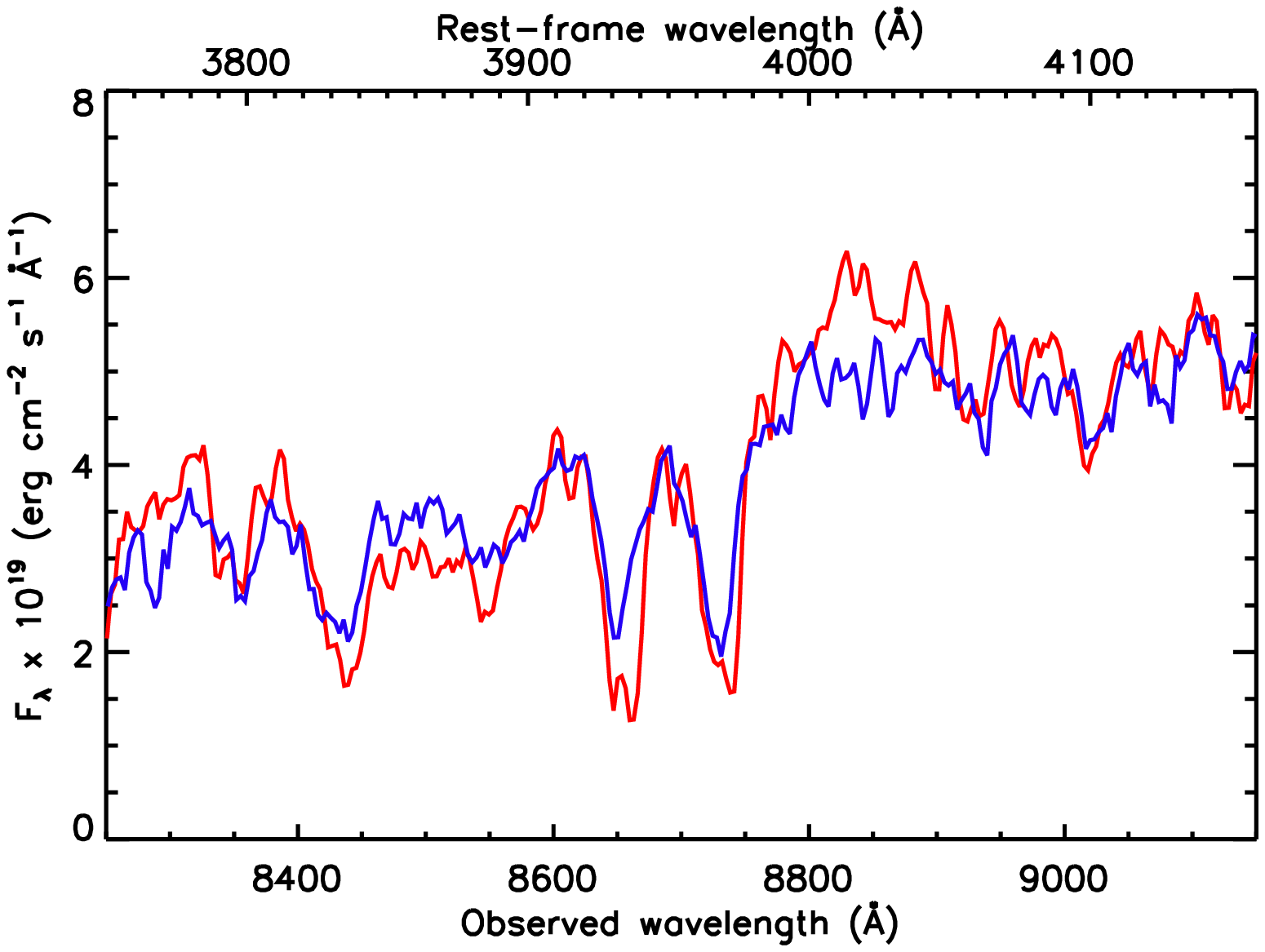}
\includegraphics[scale=0.5]{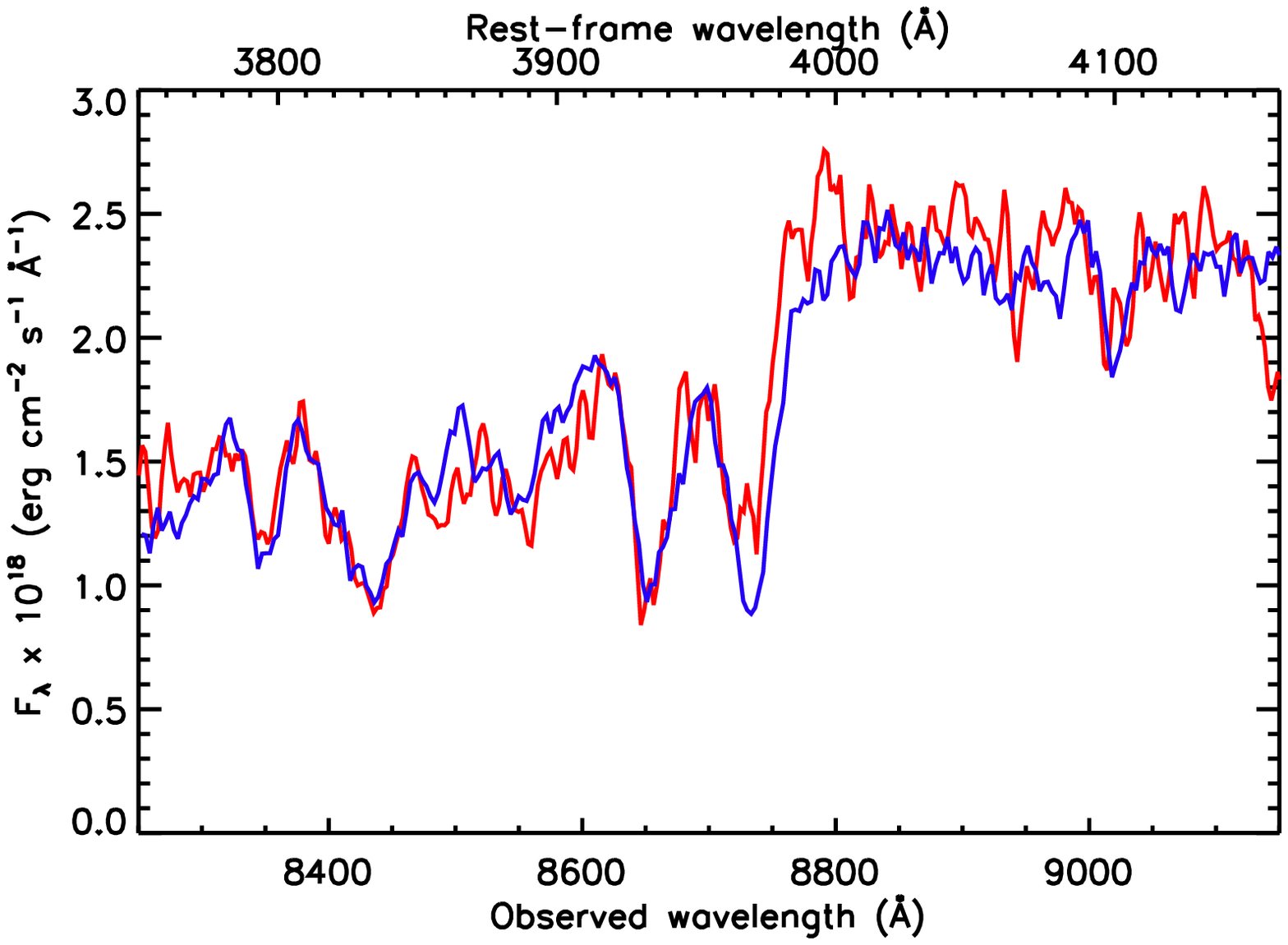}
\caption{Stacked spectra of low-mass ($M_{\star} < 1.2\times10^{11}M_{\odot}$, 
\emph{left panel}) and high-mass ($M_{\star} \geq 1.2\times10^{11}M_{\odot}$, 
\emph{right panel}) galaxies in the field (\emph{blue}) and cluster (\emph{red}) 
samples. The signal-to-noise ratios and equivalent exposure times are: 23, 47h (GOODS) 
and 17, 39h (RDCS 1252) for the low-mass bin (\emph{left}); 42, 46h (GOODS) and 13, 36h 
(RDCS 1252) for the high-mass bin (\emph{right}).}
\label{fig6}
\end{figure}

In Fig. \ref{fig4}, left panel, we plot the distribution of
the difference between the age and final formation time, $T-t_{fin}$, of the
best fit models which lie within the intersection of 3$\sigma$
confidence levels of the photometric and spectral data for our field
and cluster samples. The top axis gives the corresponding final formation
redshift. We find that the mean final formation time of the cluster early-type galaxy 
population is $\sim$1 Gyr greater than the corresponding population in the field, 
namely that field early-types have longer star formation time scales. We also note 
that the star-formation weighted ages of the two populations, as defined in Menci et 
al. (\cite{Menci08}) and Rettura et al. (\cite{Rettura08}), differ by only $\sim\! 0.5$ 
Gyr on average, as shown in Fig. \ref{fig5}. This age difference is in very good 
agreement with that derived by van Dokkum \& van der Marel (\cite{vanDokkum07}) from 
the evolution of the mass-to-light ratio, which is based on a completely independent 
method and data set. In Fig. \ref{fig4}, right panel, we plot the distribution of 
$T-t_{fin}$ of the models that lie within the 3$\sigma$ confidence level from the 
photometric data only. We see that fitting only the stacked SEDs of our cluster and 
field sample yields very similar results and that the spectroscopic data carries most 
of the weight for the difference in final formation times. This can be appreciated in 
Fig. \ref{fig6}, where we divide our samples equally in two mass bins and show that 
the averaged spectrum of the cluster galaxies has a more pronounced 4000 \AA\ break 
than that of the field sample, specifically at lower masses. This elucidates how the 
time scale of the star formation activity in the cluster environment is shorter than in 
the field, and such a difference becomes negligible for the most massive galaxies.

\begin{figure}
\centering
\includegraphics[scale=0.5]{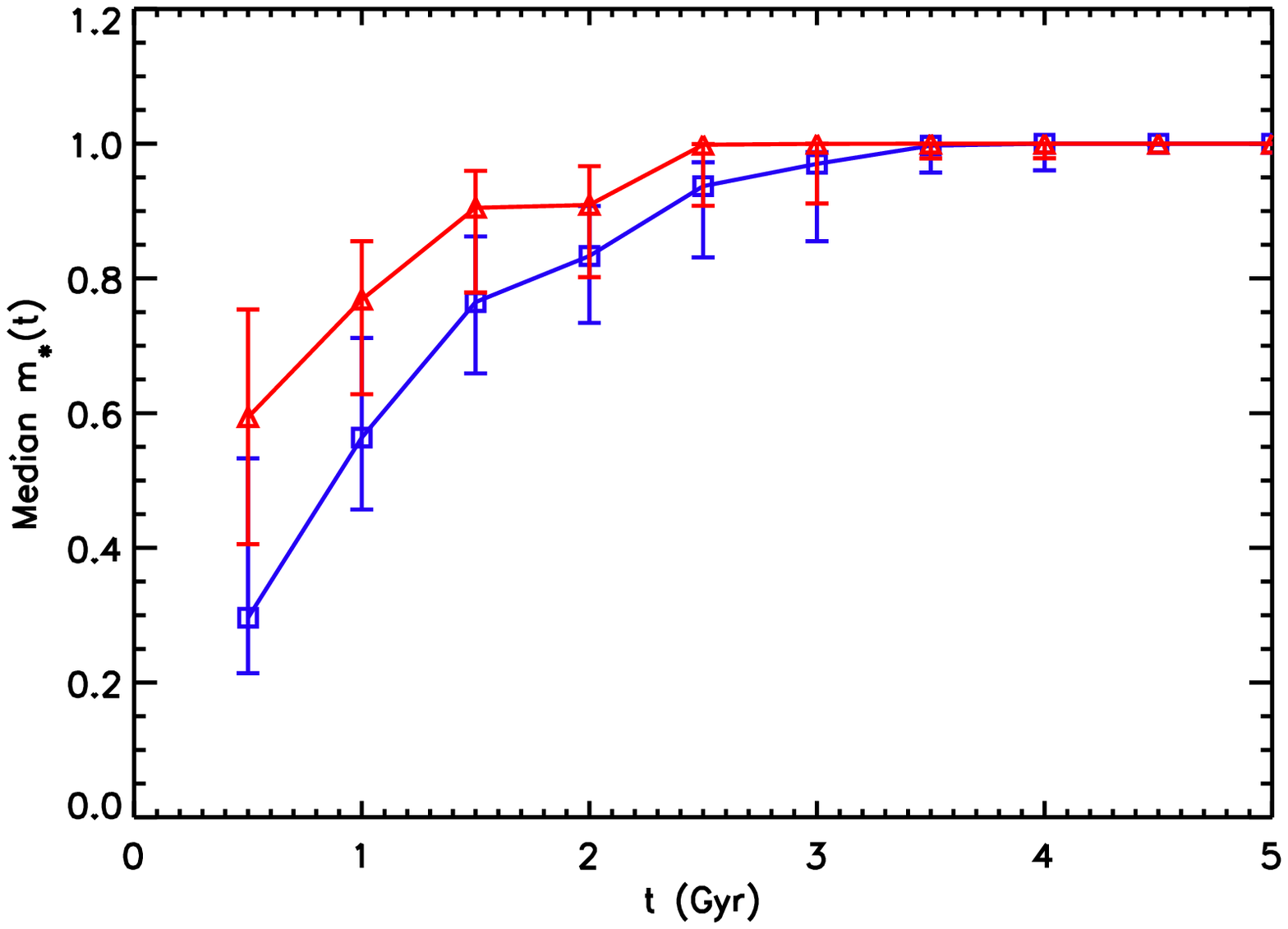}
\caption{Median stellar mass fraction of the best fit models as a function of
  the time since the onset of star formation, for the GOODS
  (\emph{blue}) and RDCS 1252 (\emph{red}) samples.}
\label{fig7}
\end{figure}

In Fig. \ref{fig7}, we plot the median stellar mass fraction,
$m_{\star}(t)$, formed at time $t$ after the onset of star
formation in cluster and field galaxies. 
This is computed as the median of the integrated star formation rate (eq.\ref{eq1}) 
of all models within the 3$\sigma$ confidence levels. The error
bars represent the standard deviation of the distribution of
$m_{\star}(t)$. This difference between the star formation histories
of our cluster and field sample is even greater if we consider only
the simple delayed exponential models.

To assess the constraining power of the spectrophotometric fit when
using low signal-to-noise spectra, we perform a large set of
Monte-Carlo simulations. We first use the BC03 model corresponding
to the best fit parameters of the stacked spectrophotometry for each
sample; for the GOODS sample we find \{$T=4.25\,{\rm Gyr},\, \tau=0.6\, {\rm
  Gyr},t_{burst}=1\,{\rm Gyr},\, A=0.5$\}, for RDCS 1252 \{$T=5\, {\rm
  Gyr},\,\tau=0.5\,{\rm Gyr},\, t_{burst}=3\,{\rm Gyr},\,
A=0.2$\}. We add to the model SED (obtained by integrating the model
spectrum over the filter curves) and spectrum an amount of noise
consistent with that of our individual galaxies. We find that the
region of parameter space occupied by the models that best fit the
input spectrum overlaps completely with confidence levels of the fit
of the input SED. Therefore, such low signal-to-noise spectra do not
add more constraints on the star formation history than those given by
the broad-band photometry alone. This confirms the need to co-add
the spectrophotometric data.

\begin{figure}
\centering
\includegraphics[scale=0.5]{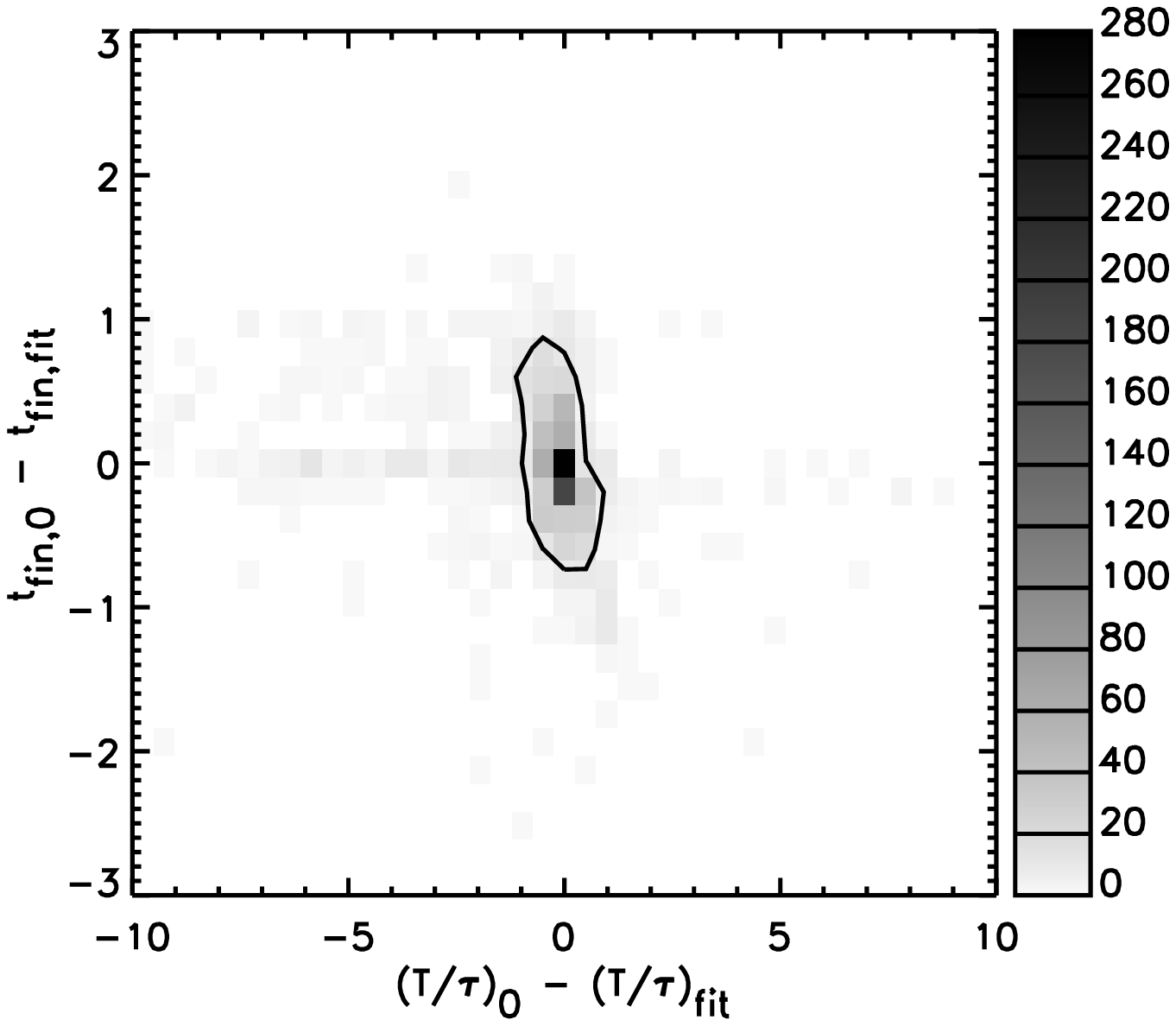}
\includegraphics[scale=0.5]{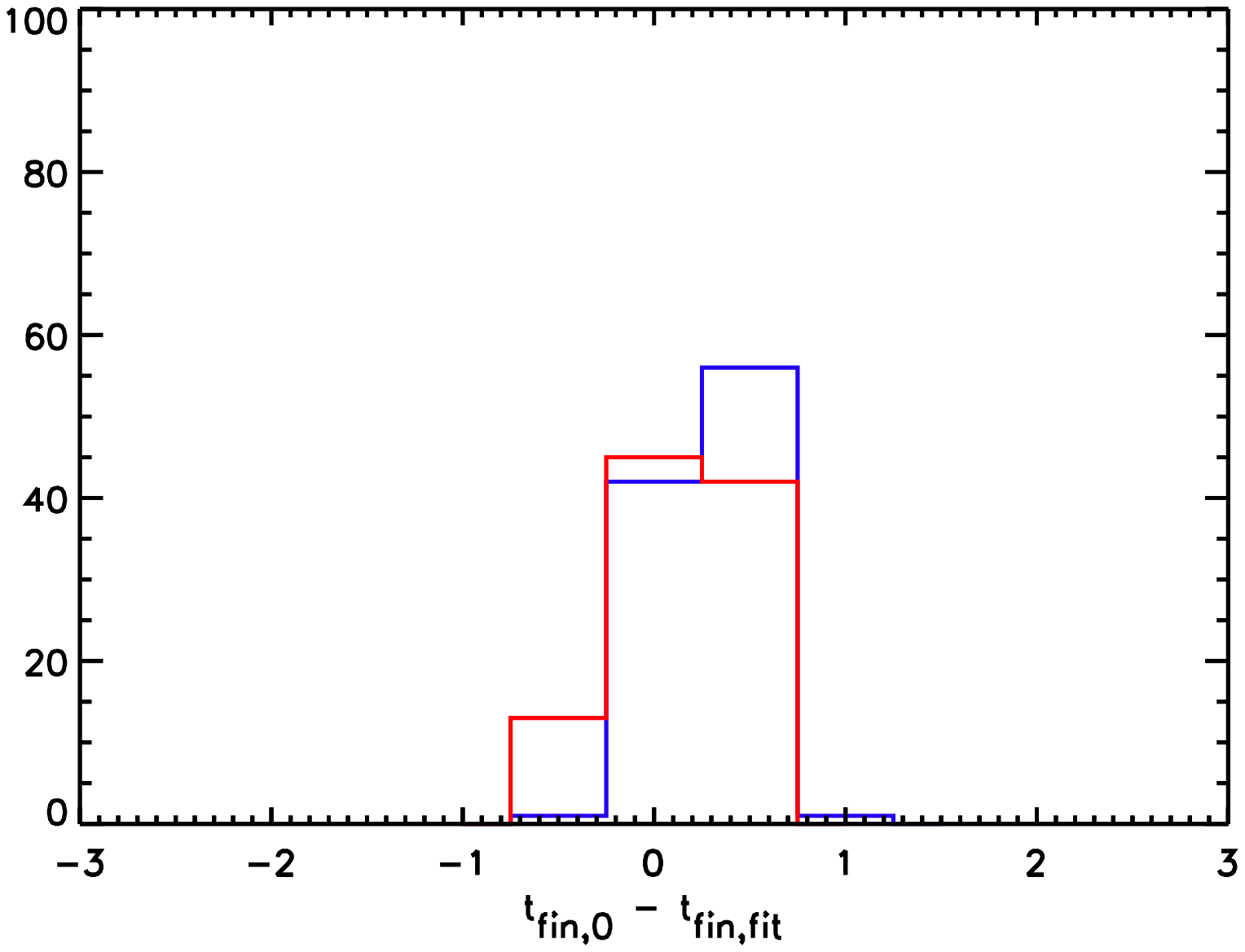}
\caption{\emph{Left panel}: distribution of the difference between the
  input and best fit solution over the grid of models
  \{$T,\tau,t_{burst},A$\} for two relevant parameters: the formation
  time $t_{fin}$ and the age/$\tau$ ratio. The gray scale refers to 
  the number of models in the grid. \emph{Right panel}:
  distribution of the residuals of final formation times obtained by
  perturbing the best fit model of the GOODS (\emph{blue}) and
  RDCS 1252 (\emph{red}) samples.}
\label{fig8}
\end{figure}

We then investigate possible biases in our fitting approach with two
different sets of simulations. First, each model spectrum in the
\{$T,\tau,t_{burst},A$\} grid is redshifted to the observed frame and
synthetic magnitudes are computed for our broad-band filters, assuming errors similar 
to those of our stacked spectrophotometric data.
The resulting SED and spectrum is fitted
using the approach described above. We then compare the parameters of
the best fitting models with those of the input model. Because of the
known degeneracy between age and $\tau$ (e.g., Gavazzi et al. 
\cite{Gavazzi02}), we do not consider them separately but rather use the
$T/\tau$ ratio. We also compare the median final formation time
$t_{fin,fit}$ of the best fit models within the intersection of the $3\sigma$
confidence levels (see Fig. \ref{fig2}) to the formation
time $t_{fin,0}$ of the input model.
The distribution of $t_{fin,0}-t_{fin,fit}$ is found to be strongly peaked
at 0 with a standard deviation of 0.5 Gyr. The distribution of
$(T/\tau)_0-(T/\tau)_{fit}$ likewise shows no bias
(Fig. \ref{fig8}, left panel). We can therefore conclude that our fit
reproduces reasonably well the original data and that any difference
we find between our field and cluster samples is not an artifact of
the method.

Secondly, we carry out a Monte-Carlo simulation on the single best fit
model for the cluster and the field sample. We add to the model SEDs
and spectra an amount of noise equal to that of our stacked
spectrophotometric data and perform the $\chi^2$ goodness-of-fit test
described above. As shown in Fig. \ref{fig8} (right panel), we
do not find any appreciable differential bias between the two samples.

To determine whether the difference between our cluster and field
samples is due to a selection effect, we also consider pairs of randomized
samples constructed from both field and cluster galaxies.
We first convert the $b_{435}$, $v_{606}$ and J magnitudes of the
GOODS sample to B, V and J$_{s}$, as used in RDCS 1252. We integrate
the spectra of our grid of models to derive the photometric
transformations at $z \simeq 1.2$ :

\begin{eqnarray}B &=& b_{435} - 0.021\times(b_{435}-v_{606}) + 0.0014\\
V &=& v_{606} + 0.32\times(b_{435}-v_{606}) + 0.081\\
J_{s} &=& J - 0.03\times(J-K_{s}) + 0.013\end{eqnarray}

In this test, we ignore the R band data of RDCS 1252 as there is no comparable 
band in the GOODS photometry.
We construct one hundred sample pairs by drawing 21 galaxies randomly
from GOODS and RDCS 1252 to constitute the 'cluster' sample and
assigning the rest to the 'field' sample. We then perform the same
fitting procedure, as described in Section \ref{sec:method}, on these two 
separate pseudo-samples.  As a measure
of the difference between the two pseudo-populations, we consider the
absolute difference in mean final formation times $\Delta<t_{fin}>$.

\begin{figure}
\centering
\includegraphics[scale=0.5]{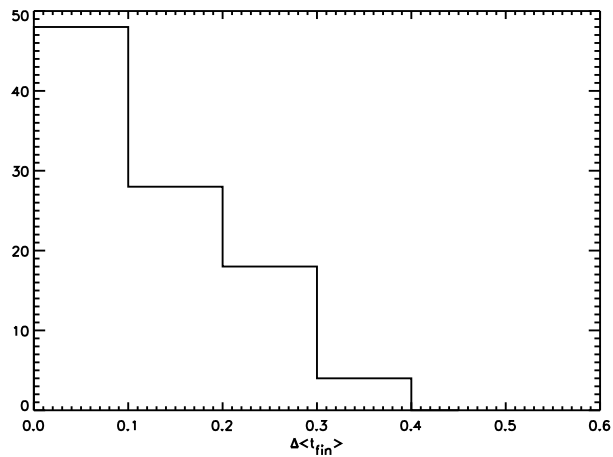}
\caption{Distribution of the difference between the mean formation
  times of 100 pairs of randomized samples.}
\label{fig9}
\end{figure}

Fig. \ref{fig9} shows the distribution of $\Delta<t_{fin}>$ in
our pseudo-cluster and field sample pairs. We see that it is peaked at
$\Delta<t_{fin}>=0$ and that the value of $\sim\! 1$ Gyr derived from the
GOODS and RDCS 1252 samples lies well outside this distribution. This
implies that the galaxies in our field and cluster samples are drawn
from two distinct populations.

We stress that our field sample is more
deficient in lower mass galaxies with respect to the total photometric
sample than our cluster sample (see Fig. \ref{fig1}). This
however does not affect our conclusions. Even if the field sample were
corrected for completeness, the averaged SED and spectrum would be expected to appear
bluer and thus younger, amplifying the difference between the star
formation histories of the field and the cluster. As widely reported
in the literature, we also find that less massive galaxies are best
fitted with younger stellar population models (the so-called ``downsizing''). 
Unfortunately, we do not have enough statistics to study whether at this redshift the
mass-age correlation varies form cluster to field.

Since age and metallicity have a similar effect on the spectrum of a
stellar population, the difference we observe between our field and
cluster samples might not correspond to distinct star formation
histories, but to a difference in metal content. To quantify this
effect, we compare our average field galaxy data to sub-solar
metallicity models computed using the GALAXEV code and the same grid
of parameters as before, keeping the cluster galaxies at solar metallicity.

\begin{figure}
\centering
\includegraphics[scale=0.5]{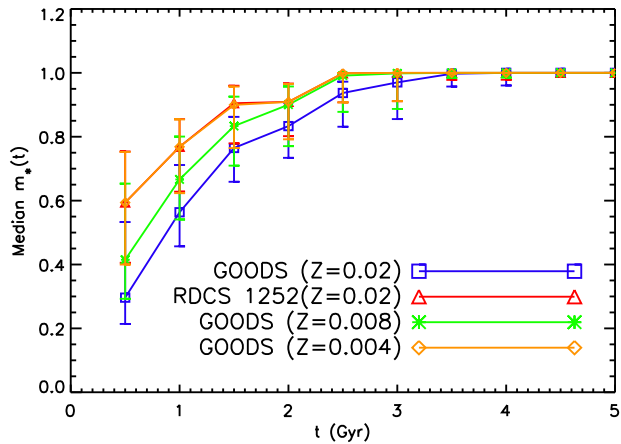}
\caption{Median stellar mass fraction of the best fit models as a
  function of the time since the onset of star formation, for RDCS 1252
  at solar metallicity and GOODS at solar and sub-solar
  ([Z/H]=0.008,0.004) metallicities.}
\label{fig10}
\end{figure}

As shown in Fig. \ref{fig10}, the star formation histories of our
cluster and field samples coincide if one population is assumed to be
approximately a factor of five more metal rich than the other. 
However, this ad hoc
assumption does not appear to be realistic. At low redshift, the
variation in metallicity of early type galaxies with environment was
found to be less than 0.1 dex (Thomas, Maraston \& Bender 
\cite{Thomas05}), with field ellipticals being more metal-rich than
their cluster counterparts. As the bulk of star formation in cluster
and field ellipticals is understood to have occurred at
$z > 2$ (Renzini \cite{Renzini06}), it is unlikely that the
relative metallicity of field and cluster ellipticals is different at
these redshifts.

As with metallicity, age and dust content are largely degenerate when
fitting SEDs. However, since we use high-resolution spectra as well as
SEDs, our method is less sensitive to dust than to metallicity.  In
this work, extinction by dust is not taken into account, as Rettura et
al. (\cite{Rettura06}) found that the SEDs of RDCS 1252.9-2927 member
galaxies are best fitted by models with no or little dust
extinction. Furthermore, there is no evidence that dust structures in
early type galaxies contribute significantly to their mid-IR spectrum
(Temi et al. \cite{Temi05}) or are correlated with environment (Tran
et al. \cite{Tran01}).

\begin{figure}
\centering
\includegraphics[scale=0.5]{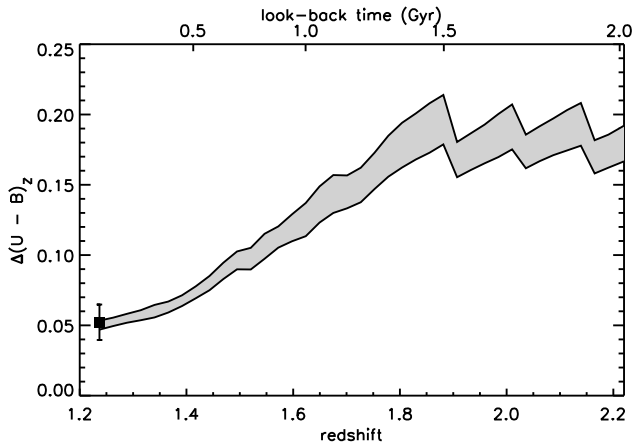}
\caption{Predicted rest-frame $(U-B)_z$ scatter of the red sequence of RDCS 
1252 extrapolated to $z=2.2$. The shaded area represents the rms dispersion 
of the models that best fit the SEDs of all red sequence galaxies. The 
filled square is derived from the $i_{775}-z_{850}$ scatter measured by 
Blakeslee et al. (\cite{Blakeslee03}).}
\label{fig11}
\end{figure}

We also measure the synthetic scatter of the rest-frame $(U-B)_z$
color, $\Delta(U-B)_z$, of the early-type (E and S0) galaxies lying on
the red sequence of RDCS 1252 using the morphological classification
and selection criterion described by Blakeslee et
al. (\cite{Blakeslee03}), for galaxies brighter than $z_{850}=24.0$
(this yields 33 red sequence galaxies).  Rest frame $(U-B)_z$
colors are obtained for each galaxy using the model that best fits the
9-band SED. For this purpose, we use only simple delayed,
exponentially declining star formation histories parametrized by $T$
and $\tau$ (see Section \ref{sec:method}).  We find good agreement
with the Blakeslee et al. measurement (0.05 mag intrinsic scatter). 
We can then use these models to predict the evolution of the $(U-B)_z$ scatter 
with redshift and corresponding look-back time from $z=1.24$ up to $z=2.2$.
At each redshift, the scatter is measured applying a 3$\sigma$ clipping around 
the linear best fit, as customary in the literature. A few additional galaxies 
drop out of the red sequence sample as the redshift becomes larger than their 
formation redshift. We estimate the uncertainty on the scatter at each
redshift by randomly perturbing the SED of each galaxy, assuming
Gaussian errors on the fluxes.  Fig. \ref{fig11} shows that the
red sequence dissipates at $z\sim 1.9$, reaching scatter values comparable 
with those observed in the red population of the field up to $z\sim2$ 
(Cassata et al. \cite{Cassata08}). At $z\sim1.9$, 25 galaxies out of the 
original 33 are still included in the red sequence sample. This redshift is 
also consistent with the median final formation redshift derived
from the analysis of the stacked spectrophotometric data. A similar analysis
using $(U-V)_z$ and $(U-R)_z$ scatters yields the same result.
We can conclude that the red sequence in such a massive
cluster is established in $\sim\! 1$ Gyr, and it is therefore
not surprising that recent analysis of forming red sequences in
protoclusters at $z>2$ find a significant scatter (Kodama et
al. \cite{Kodama07}, Tanaka et al. \cite{Tanaka07}, Zirm et al. \cite{Zirm08}).


\section{Conclusion}

We compared the underlying stellar population properties of $z \simeq
1.2$ early-type galaxies in the very high density environment of the
massive cluster RDCS 1252 with those in the GOODS field.  We derived
star formation histories by fitting both the spectroscopic and the
broad band photometric data with a large grid of stellar population
synthesis models.  To this purpose, we select 43 early-type galaxies
(21 in the field and 22 in the cluster) all with stellar masses greater
than $M_{lim}=5\times10^{10}M_{\odot}$, $i_{775}-z_{850} \geq 0.8$ and the
absence of the [OII]$\lambda$3727 line in the spectra. These criteria naturally 
select all galaxies on the red sequence of RDCS 1252. For each sample, we 
use the co-added spectrophotometric data of the galaxies and compare them with 
BC03 models of exponentially declining star formation rates with an 
additional burst.

We find a small although significant difference in the star formation
histories of the cluster and field populations, suggesting that the
cluster galaxies form the bulk of their stars $\sim\!0.5$ Gyr earlier
than their counterparts in the field, with massive early-type galaxies
having already finished forming stars at $z>1.5$ in both environments. 
This difference is particularly evident at masses $\lesssim 10^{11} 
M_\odot$, which are characterized by a longer star formation time scale, 
resulting in a final formation time delayed by $\sim$1 Gyr, whereas it 
becomes negligible for the most massive galaxies. 
While our differential analysis of the stellar population 
parameters of cluster and field galaxies in the same mass range convincingly 
shows distinct star formation histories, the absolute age difference remains 
model dependent. In an accompanying paper, Rettura et al. (\cite{Rettura08}) 
have analyzed the rest-frame far-UV flux of the same sample of early-type 
galaxies and found that field galaxies are at least 0.5 mag brighter than the 
RDCS 1252 galaxies in the same mass range. Our best fit models consistently 
predict this magnitude difference, which is indicative of the longer star 
formation time scale of the field galaxies. 
We have verified that such a difference in derived star formation
histories in the two environments cannot be ascribed to incompleteness
of the mass selected samples, which would tend to rather increase such
an effect.  We also used extensive Monte-Carlo simulations to identify
possible biases in the model fit, and discussed the effect of inherent
degeneracies such as metallicity and dust. We note that independent
studies of massive early-type galaxies based on the measurement of the
mass-to-light ratios of massive early-type galaxies in high- and
low-density environments (Treu et al. \cite{Treu05}, van der Wel et
al. \cite{vanderWel05}, van Dokkum \& van der Marel
\cite{vanDokkum07}) have reached similar conclusions.

We also used the best fit star formation histories from the 9-band
SEDs of the red sequence galaxies in RDCS 1252 to predict that a tight
($\Delta(U-B)_z=0.05$ mag) red sequence at $z=1.2$ is established over
approximately 1 Gyr and dissolves by $z\approx 1.9$. This suggests that for
massive clusters, which have long reached virialization by redshift
1.2, we do not expect a significant red sequence at $z>2$, i.e. a $(U-B)_z$ 
color scatter well above 0.1 mag.

These observations and analysis can be used to provide significant
constraints on galaxy evolution models in a hierarchical scenario
(e.g. Menci et al. \cite{Menci08}), which predicts the evolution in
high-density environments to be accelerated compared to the field.

\begin{acknowledgements}
This research was partially supported by the DFG cluster of excellence 
``Origin and Structure of the Universe'' (www.universe-cluster.de)
\end{acknowledgements}


\begin{thebibliography}{}

\bibitem[2003]{Blakeslee03}{Blakeslee, J.P., Franx, M., Postman, et al. 2003, \apj, 596, L143}

\bibitem[2003]{BC03}{Bruzual, G., Charlot, S. 2003, \mnras, 344, 1000}

\bibitem[2000]{Benitez00}{Ben\'{i}tez, N. 2000, \apj, 536, 571}

\bibitem[1998]{Bernardi98}{Bernardi, M., Renzini, A., Da Costa, L.N. et al. 1998, \apj, 507, L43}

\bibitem[2008]{Cassata08}{Cassata, P., Cimatti, A., Kurk, J. et al. 2008, \aap, 483, 39}

\bibitem[2006]{Clemens06}{Clemens, M.S., Bressan, A., Nikolic, B. et al. 2006, \mnras, 370, 702}

\bibitem[1980]{Coleman80}{Coleman, G.D., Wu, C.-C., Weedman, D.W. 1980, \apjs, 43, 393}

\bibitem[2006]{DeLucia06}{De Lucia, G., Springel, V., White, S.D.M., Kauffmann, G. 2006, \mnras, 366, 499}

\bibitem[2007]{Demarco07}{Demarco, R., Rosati, P., Lidman, C. et al. 2007, \apj, 663, 164D}

\bibitem[2006]{Alighieri06}{di Serego Alighieri, S., Lanzoni, B., J\o rgensen, I. 2006, \apj, 647, L99}

\bibitem[1962]{Eggen62}{Eggen, O.J., Lynden-Bell, D., Sandage, A.R. 1962, \apj, 136, 748}

\bibitem[1996]{Gavazzi96}{Gavazzi, G., Pierini, D., Boselli, A. 1996, \aap, 312, 397}

\bibitem[2002]{Gavazzi02}{Gavazzi, G., Bonfanti, C., Sanvito, G., Boselli, A., Scodeggio, M. 2002, \apj, 576, 135}

\bibitem[2004]{Giavalisco04}{Giavalisco, M., Ferguson, H.C., Koekemoer, A.M. et al. 2004, \apj, 600, 93}

\bibitem[2007]{Haeussler07}{H\"{a}ussler, B., McIntosh, D.H., Barden, M. et al. 2007, \apjs, 172, 615}

\bibitem[2007]{Kodama07}{Kodama, T., Tanaka, I., Kajisawa, M. et al. 2007, \mnras, 377, 1717}

\bibitem[2004]{Lidman04}{Lidman, C., Rosati, P., Demarco, R. et al. 2004, \aap, 416, 829}

\bibitem[2008]{Menci08}{Menci, N., Rosati, P., Gobat, R. et al. 2008, \apj, in press}

\bibitem[1974]{Oke74}{Oke, J.B. 1974, \apjs, 27, 21}

\bibitem[1997]{Poggianti97}{Poggianti, B.M., Barbaro, G. 1997, \aap, 325, 1025}

\bibitem[2006]{Renzini06}{Renzini, A. 2006, \araa, 44, 141}

\bibitem[2006]{Rettura06}{Rettura, A., Rosati, P., Strazzullo, V. et al. 2006, \aap, 458, 717}

\bibitem[2008]{Rettura08}{Rettura et al., 2008, submitted to \apj}

\bibitem[2004]{Rosati04}{Rosati, P., Tozzi, P., Ettori, S. et al. 2004, \apj, 127, 230}

\bibitem[1955]{Salp55}{Salpeter, E.E. 1955, \apj, 121, 161}

\bibitem[2006]{Sanchez-Blazquez06}{S\'{a}nchez-Bl\'{a}zquez, P., Gorgas, J., Cardiel, N., Gonz\'{a}lez, J.J. 2006, \aap, 457, 809}

\bibitem[1986]{Sandage86}{Sandage, A. 1986, \aap, 161, 89}

\bibitem[2006]{Strazzullo06}{Strazzullo, V., Rosati, P., Stanford, S.A. et al. 2006, \aap, 450, 909}

\bibitem[2007]{Tanaka07}{Tanaka, M., Tadayuki, K., Kajisawa et al. 2007, \mnras, 377, 1206}

\bibitem[2005]{Temi05}{Temi, P., Brighenti, F., Mathews, W.G. 2005, \apj, 635, L25}

\bibitem[2005]{Thomas05}{Thomas, D., Maraston, C., Bender, R., Mendes de Oliveira, C. 2005, \apj, 621, 673}

\bibitem[2004]{Toft04}{Toft, S., Mainieri, V., Rosati, P. et al. 2004, \aap, 422, 29}

\bibitem[1977]{Toomre77}{Toomre, A. 1977, In \emph{Evolution of Galaxies and Stellar Populations}, ed. BM Tinsley \& RB Larson, New Haven:Yale University Observatory, p. 401}

\bibitem[2001]{Tran01}{Tran, H.D., Tsvetanov, Z., Ford, H.C., Davies, J. 2001, \apj, 121, 2928}

\bibitem[2005]{Treu05}{Treu, T., Ellis, R.S., Liao, T.X., van Dokkum, P.G. 2005, \apj, 622, 5}

\bibitem[2005]{vanderWel05}{van der Wel, A., Franx, M., van Dokkum, P.G., Rix, H.-W., Illingworth, G.D., Rosati, P. 2005, \apj, 631, 145}

\bibitem[2001]{vanDokkum01}{van Dokkum, P.G., Franx, M., Kelson, D.D., Illingworth, G.D. 2001, \apj, 553, L39}

\bibitem[2007]{vanDokkum07}{van Dokkum, P.G., van der Marel, R.P. 2007, \apj, 655, 30}

\bibitem[2006]{Vanzella06}{Vanzella, E., Cristiani, S., Dickinson, M. et al. 2006, \aap, 454, 423}

\bibitem[2008]{Vanzella08}{Vanzella, E., Cristiani, S., Dickinson, M. et al. 2008, \aap, 478, 83}

\bibitem[2007]{Zirm08}{Zirm et al. 2008, \apj, 680, 224}

\end{thebibliography}
\end{document}